\documentclass[12pt]{article}

\usepackage{amssymb}
\usepackage{bm}
\usepackage[usenames]{color}
\usepackage{arydshln}
\usepackage{amsmath}
\usepackage{color}
\usepackage{graphics}
\usepackage{amsthm}
\usepackage{ascmac}
\usepackage{amsmath}
\usepackage{amsfonts}
\usepackage{amssymb}
\usepackage{multicol}
\usepackage{epsfig}
\usepackage{here}
\usepackage{authblk}

\usepackage{setspace}
\usepackage{listings}
\usepackage{ulem}
\makeatletter

\@addtoreset{equation}{section}
\makeatother

\newtheorem{theo}{Theorem}

\newtheorem{assu}{Assumption}

\newtheorem{prop}{Proposition}

\newcommand{\indep}{\mathop{\perp\!\!\!\perp}}

\newcommand{\bld}{\boldsymbol}

\newcommand{\argmax}{\mathop{\rm arg~max}\limits}

\usepackage[top=1in,bottom=1in,left=1in,right=1in]{geometry}

\title{Likelihood-based Instrumental Variable Methods for Cox Proportional Hazard Models}
\author[1]{Shunichiro Orihara}

\affil[1]{Graduate School of Data Science, Yokohama City University, Kanagawa, Japan}

\date{}

\begin{document}
\allowdisplaybreaks[4]
\begin{singlespace}
\maketitle
\end{singlespace}
\section*{Abstract}
In biometrics and related fields, the Cox proportional hazards model are widely used to analyze with covariate adjustment. However, when some covariates are not observed, an unbiased estimator usually cannot be obtained. Even if there are some unmeasured covariates, instrumental variable methods can be applied under some assumptions. In this paper, we propose the new instrumental variable estimator for the Cox proportional hazards model. The estimator is the similar feature as Mart\'{i}nez-Camblor et al., 2019, but not the same exactly; we use an idea of limited-information maximum likelihood. We show that the estimator has good theoretical properties. Also, we confirm properties of our method and previous methods through simulations datasets.

\vspace{0.5cm}
\noindent
{\bf Keywords}: Causal inference, Cox proportional hazard model, EM algorithm, Instrumental variable, Limited-information maximum likelihood, Probit model, Unmeasured covariates
\newpage
\section{Introduction}
In biometrics and related fields, we commonly encounter time-to-event outcomes. The Cox proportional hazards model (CPHM, Cox, 1972) are widely used to analyze with covariate (or confounders; hereafter, we call ``covariates") adjustment for outcomes since the hazard ratio is one of the useful descriptive statistics under the proportional hazard assumption. When all covariates are observed, the covariates can be adjusted and an unbiased estimator for causal effects can be obtained; the situation of ``no unmeasured confounding" (c.f. Hern\'{a}n and Robins, 2020). No unmeasured confounding is one of the sufficient assumptions to estimate an unbiased estimator for causal effects. Whereas, when some covariates are not observed, an unbiased estimator usually cannot be obtained (Gail et al., 1984); the situations where there are some unmeasured covariates. Unmeasured covariates constitute one of the important problems in causal inference since no unmeasured confounding is no longer held. Therefore, a different sufficient assumption needs to be applied. In this paper, we focus on instrumental variable (IV) methods. Although, there are some theoretical results and applications in recent years (Pierce et al., 2011, Baiocchi et al., 2014, Kang et al., 2016, Burgess et al., 2017, Guo et al., 2018, Windmeijer et al., 2019, and Orihara, 2021a), ordinary IV methods such as two-stage least square (2SLS) estimators assume continuous outcomes; we cannot apply the IV methods not only to time-to-event outcomes but also dichotomous outcomes simply.

For dichotomous outcomes, Terza et al. (2008) introduced a two-stage residual inclusion (2SRI) estimator similar to the control function approach (Wooldridge, 2010). 2SRI is another two-step procedure expanded to include nonlinear models, such as logistic regression and probit models, whereby an unbiased estimate of the causal effect can be obtained even when there are nonlinear models. Although 2SRI overcomes the problem of 2SLS, it may derive biased causal effects, as mentioned in Basu et al. (2017) and Wan et al. (2018). Also, since 2SRI is necessary to estimate the residuals of a treatment variable, continuous outcomes are commonly assumed. According to the simulation results of Basu et al. (2017), a full-likelihood approach derives a more accurate estimate than 2SRI (see also Section 5 of Burgess et al., 2017). A limited-information maximum likelihood (LIML) estimator (Wooldridge, 2014) uses a full-likelihood approach, but has features similar to those of 2SRI and the control function approach. In Orihara (2021b), performance of 2SRI and LIML is confirmed under some simulation settings, and it is mentioned that LIML can estimate causal effects more accurate than 2SRI with model selection procedures.

For time-to-event outcomes, some methods have been proposed in recent years. Tchetgen Tchetgen et al., 2015 and Ying et al., 2019 assume the additive hazard model, and apply the control function approach and a 2SRI, respectively. Kianian et al., 2019, Mart\'{i}nez-Camblor et al., 2019, Cui et al., 2020, and Mart\'{i}nez-Camblor et al., 2021 assume the CPMH, considered mainly in this paper. Kianian et al., 2019 extend the weighting method proposed by Abadie, 2003 to construct an estimator of the hazard ratio, and consider properties of the proposed estimator. Since their method is a direct extension of Abadie, 2003, only a binary and univariate IV can be applied. Mart\'{i}nez-Camblor et al., 2019 apply the 2SRI approach, and uses an idea of the frailty model (Nielsen et al., 1992); this is because an unobserved variability is considered as a ``frailty". They derive useful properties of the proposed estimator under a continuous treatment situation. Cui et al., 2020 and Mart\'{i}nez-Camblor et al., 2021 propose estimators for the marginal hazard ratio (Hern\'{a}n et al., 2000) under a time-varying treatment situation and only a point treatment situation, respectively. Although their methods are restricted to a binary or univariate IV and situation where unmeasured covariates have a strong independence, the marginal hazard ratio can be interpreted as the causal hazard ratio (c.f. Hern\'{a}n and Robins, 2020).

As mentioned previously, there are many proposals when applying IV methods to CPMH. However, to best of our knowledge, no methods have been proposed for more general situations; more precisely, we need to assume either ``a binary and univariate IV", ``a continuous treatment", or ``a strong independence of unmeasured covariates" for almost all methods. In this paper, we would like to extend their results to more broad situations: considering both a binary treatment (or a continuous treatment), and non-binary and multiple IVs.  These extensions are important contributions since a binary treatment situation is commonly assumed in biometrics and related fields, and non-binary and multiple IVs are commonly used for the Mendelian randomization (c.f. Burgess et al., 2017). Concretely, we consider the similar approach as Mart\'{i}nez-Camblor et al., 2019, but not the same exactly; we use an idea of the frailty model directly to capture the variability of unmeasured covariates. Also, by using an idea of LIML, treatment models can be flexible enough to consider since the residuals for treatment values need not be considered; only a likelihood is necessary. From these idea, theoretical properties of our proposed method can be proved simply (Zeng et al., 2005 and Gamst et al., 2009), and the estimating procedures for frailty models can be applied for our proposed method (e.g. Vaida and Xu, 2000): we can construct EM algorithm (Dempster et al., 1977).

The remainder of the paper proceeds as follows. In section 2, we will introduce a motivation example, and present a model considered in this paper. In section 3, we derive an observed likelihood that is integrated by unobserved covariates conditional on observed data, and constructing E-step and M-step of the EM algorithm. By following the proof of Zeng et al., 2005 and Gamst et al., 2009, we proof the consistency and the asymptotic normality of estimated parameters. In section 4, we confirm properties of our method through some simulation settings. All regularity conditions, calculations, and proofs are given in appendix.

\section{Notations}

Let $n$ be the sample size, and assume that $i = 1,\ 2,\dots, n$ are i.i.d. samples. $\bld{X}\in\mathcal{X}\subset\mathbb{R}^{p}$ and $\bld{Z}\in\mathcal{Z}\subset\mathbb{R}^{K}$ denote a vector of covariates and a vector of IVs respectively. In the context of instrumental variables methods, observed covariates $\bld{X}$ are sometimes not considered. However, to consider more general situations, the observed covariates are included in the following discussions. The following relationship is assumed for the unmeasured variables:
\begin{align}
\label{for1_1}
\left(
\begin{array}{c}
V\\
U
\end{array}
\right)\sim N_{2}\left(\bld{0}_{2},\left(
\begin{array}{cc}
1&\rho\sigma_{u}\\
&\sigma_{u}^2
\end{array}
\right)\right),\ \ \left(
\begin{array}{c}
V\\
U
\end{array}
\right)\indep\left(
\begin{array}{c}
\bld{X}\\
\bld{Z}
\end{array}
\right).
\end{align}
Also, we introduce models for a treatment value $W\in\mathcal{W}\subset\mathbb{R}$ considered in this paper:
\begin{align}
\label{for1_2}
W=\bld{1}\left\{\tilde{\bld{X}}_{w}^{\top}\bld{\alpha}+V\geq 0\right\},
\end{align}
where ${\rm E}[V]=0,\ Var(V)<\infty$. In other words, $V$ is an error or latent variable related to the treatment $W$; a likelihood of the treatment $W$ is described as $f(w|\bld{z},\bld{x};\bld{\alpha})$, where $f$ is a probability function. Since $V$ is the standard normal distribution, (\ref{for1_2}) is a probit model. As mentioned in the future works, we need to extend the distribution to more broad distribution class such as a logistic regression model. Note that $\tilde{\bld{x}}_{w}$ includes an intercept, $\bld{z}$, and $\bld{x}$; for instance
$$
\tilde{\bld{x}}_{w}^{\top}\bld{\alpha}=\left(1,\bld{z}^{\top},\bld{x}^{\top}\right)\left(
\begin{array}{c}
\alpha_{0}\\
\bld{\alpha}_{z}\\
\bld{\alpha}_{x}
\end{array}
\right).
$$
Note also that the same discussions described hereinafter are hold for continuous treatment models. However, to simplify the following discussion, dichotomous treatment is only considered in this manuscript.

Next we introduce an outcome model. In this paper, right-censored situations are considered. Let $\tilde{T}\in\mathcal{T}\subset\mathbb{R}$ and $C\in\mathcal{T}\subset\mathbb{R}$ denote the event time and the censoring time, respectively. Also, we assume that there are no tie data. The observed time denote $T=\min(\tilde{T},C)$ and $\delta$ denote the indicator of the censoring; $\delta=\bld{1}(T\leq C)$. Under these settings, we consider the following frailty model:
\begin{align}
\label{for1_4}
\lambda(t)=\lambda_{0}(t)\exp\left\{\tilde{\bld{x}}_{t}^{\top}\bld{\beta}+u\right\},
\end{align}
where $\lambda_{0}(t)$ indicates the baseline hazard. In this paper, the unmeasured variable $U$ is called as ``unmeasured covariates". Under this formulation, an unmeasured covariate problem is occurred when estimating some treatment effects since there are relationship not adjusted sufficiently between the treatment variable and some outcome variables. Note that $\tilde{\bld{x}}_{t}$ includes an intercept, $w$, and $\bld{x}$; for instance
$$
\tilde{\bld{x}}_{t}^{\top}\bld{\beta}=\left(1,w,\bld{x}^{\top},w\bld{x}^{\top}\right)\left(
\begin{array}{c}
\beta_{0}\\
\beta_{w}\\
\bld{\beta}_{x}\\
\bld{\beta}_{wx}
\end{array}
\right).
$$
We are interested in the treatment effect $\beta_{w}$ and the interaction term $\bld{\beta}_{wx}$. From (\ref{for1_4}), a likelihood of the frailty model is described as 
\begin{align*}
f(t|w,\bld{x},u;\bld{\beta})&=\lambda_{0}(t)\exp\left\{\tilde{\bld{x}}_{t}^{\top}\bld{\beta}+u\right\}\exp\left\{-\Lambda(t)\exp\left\{\tilde{\bld{x}}_{t}^{\top}\bld{\beta}+u\right\}\right\},
\end{align*}
where $\Lambda(t)=\int^{t}_{0}\lambda(t')dt'$ is a cumulative hazard function. The model (\ref{for1_4}) is similar to the standard frailty model (Nielsen et al., 1992) except that the unobserved covariate $U$ is related to the the treatment $W$ and a cluster size is only one. The latter feature affects to a parameter identification directly; we explain it later. Note that regarding (\ref{for1_4}), any parametric models are not assumed to the baseline hazard. 

For time-to-event outcome, we usually take care of censoring data. To consider a censor $C$, we assume the following assumption:
\begin{assu}{$\phantom{}$}\\
\label{assu1}
Regarding the censor $C$, the following independences hold:
\begin{enumerate}
\item $T\indep C|U,W,\bld{Z},\bld{X}$
\item $C\indep U|W,\bld{Z},\bld{X}$
\end{enumerate}
There are equivalent to the following condition (c.f. Glymour et al., 2016):
$$
C\indep (T,U)|W,\bld{Z},\bld{X}
$$
\end{assu}\noindent
About {\bf Assumption \ref{assu1}}, the former form is easy applying to the following discussions, meanwhile the latter form is easy to interpret; the censor and the joint distribution of the event time and the unobserved covariate are conditionally independent given some observed variables. This is an extension of an ordinary assumption to the censor; the independency of the unmeasured covariate is necessary for the following proposition.

From the above, a joint distribution of the observed time $T$ and the indicator of the censoring $\delta$ becomes
\begin{align}
\label{for1_5}
f(t,\delta=0|w,\bld{x},u,\bld{z})&=f_{c}(t|w,\bld{x},\bld{z})S_{t|w,x,u}(t;\bld{\beta}),\nonumber\\
f(t,\delta=1|w,\bld{x},u,\bld{z})&=f(t|w,\bld{x},u;\bld{\beta})S_{c|w,x,z}(t),
\end{align}
where $f_{c}(\cdot)$, $S_{c|w,x,z}(\cdot)$, and $S_{t|w,x,u}(\cdot)$ denote the density function of the censor, the survival function of the censor, and the survival function of the event time, respectively (c.f. Klein and Moeschberger, 2006). To estimate parameter $\bld{\beta}$, we will construct an observed likelihood with excluding the effect of the unmeasured covariate $U$.

\section{Likelihood-based Estimation, Property, and EM Algorithm}
\subsection{Observed likelihood and parameter identification}
To estimate parameter $\bld{\theta}={(\bld{\alpha}^{\top},\bld{\beta}^{\top},\bld{\xi}^{\top})}^{\top}\in\Theta$, an observed log-likelihood is considered. Note that $\bld{\xi}=(\sigma_{u},\rho)^{\top}$. The joint distribution becomes
\begin{align}
\label{for2_3}
\prod_{i=1}^{n}f(t_{i},\delta_{i},w_{i},u_{i}|\bld{z}_{i},\bld{x}_{i})&=\prod_{i=1}^{n}\left(S_{c|w,x,z}(t_{i})\right)^{\delta_{i}}\left(f_{c}(t_{i}|w_{i},\bld{x}_{i},\bld{z}_{i})\right)^{1-\delta_{i}}\nonumber\\
&\hspace{0.5cm}\times\prod_{i=1}^{n}\left(f(t_{i}|w_{i},\bld{x}_{i},u_{i};\bld{\beta})\right)^{\delta_{i}}\left(S_{t|w,x,u}(t_{i};\bld{\beta})\right)^{1-\delta_{i}}\nonumber\\
&\hspace{0.5cm}\times\prod_{i=1}^{n}f(u_{i}|w_{i},\bld{z}_{i},\bld{x}_{i};\bld{\gamma})f(w_{i}|\bld{z}_{i},\bld{x}_{i};\bld{\alpha})\nonumber\\
&\propto\prod_{i=1}^{n}\left(f(t_{i}|w_{i},\bld{x}_{i},u_{i};\bld{\beta})\right)^{\delta_{i}}\left(S_{t|w,x,u}(t_{i};\bld{\beta})\right)^{1-\delta_{i}}\nonumber\\
&\hspace{0.5cm}\times\prod_{i=1}^{n}\left(\int_{-\tilde{\bld{x}}_{w,i}^{\top}\bld{\alpha}}^{\infty}f(u_{i},v;\bld{\xi})dv\right)^{w_{i}}\left(\int^{-\tilde{\bld{x}}_{w,i}^{\top}\bld{\alpha}}_{-\infty}f(u_{i},v;\bld{\xi})dv\right)^{1-w_{i}}.
\end{align}
The proportional symbol ``$\propto$" is in the sense of the relationship of the parameter $\bld{\theta}$. Then, the log-likelihood becomes
\begin{align}
\label{for2_5}
\ell(\bld{\theta}|\bld{u})&=\sum_{i=1}^{n}\delta_{i}\log\lambda(t_{i})+\sum_{i=1}^{n}\delta_{i}\tilde{\bld{x}}_{t,i}^{\top}\bld{\beta}+\sum_{i=1}^{n}\delta_{i}u_{i}-\sum_{i=1}^{n}\Lambda(t_{i})\exp\left\{\tilde{\bld{x}}_{t,i}^{\top}\bld{\beta}+u_{i}\right\}\nonumber\\
&\hspace{0.5cm}+\sum_{i=1}^{n}w_{i}\log\left\{\int_{-\tilde{\bld{x}}_{w,i}^{\top}\bld{\alpha}}^{\infty}f(u_{i},v;\bld{\xi})dv\right\}+\sum_{i=1}^{n}(1-w_{i})\log\left\{\int^{-\tilde{\bld{x}}_{w,i}^{\top}\bld{\alpha}}_{-\infty}f(u_{i},v;\bld{\xi})dv\right\}.
\end{align}
From (\ref{for2_5}), the observed log-likelihood becomes
$$
\ell^{o}(\bld{\theta})=\int_{\bld{u}}\ell(\bld{\theta}|\bld{u})d\bld{u}.
$$
Actually, to construct the nonparametric MLE $\hat{\bld{\theta}}$, it is necessary to modify (\ref{for2_5}) slightly:
\begin{align}
\label{m_ll}
\ell(\bld{\theta}|\bld{u})&=\sum_{i=1}^{n}\delta_{i}\log\Lambda\{t_{i}\}+\sum_{i=1}^{n}\delta_{i}\tilde{\bld{x}}_{t,i}^{\top}\bld{\beta}+\sum_{i=1}^{n}\delta_{i}u_{i}-\sum_{i=1}^{n}\Lambda(t_{i})\exp\left\{\tilde{\bld{x}}_{t,i}^{\top}\bld{\beta}+u_{i}\right\}\nonumber\\
&\hspace{0.5cm}+\sum_{i=1}^{n}w_{i}\log\left\{\int_{-\tilde{\bld{x}}_{w,i}^{\top}\bld{\alpha}}^{\infty}f(u_{i},v;\bld{\xi})dv\right\}+\sum_{i=1}^{n}(1-w_{i})\log\left\{\int^{-\tilde{\bld{x}}_{w,i}^{\top}\bld{\alpha}}_{-\infty}f(u_{i},v;\bld{\xi})dv\right\},
\end{align}
where $\Lambda\{t\}$ is the jump size of $\Lambda(t)$ at $t$. From (\ref{m_ll}) with integrating out of $\bld{u}$, the nonparametric MLE $\hat{\bld{\theta}}$ can be obtained; specific construction of the observed log-likelihood of (\ref{m_ll}) will be explained later.

These formulations are the special case of Gamst et al., 2009; these are regarded as the frailty model with $n$ clusters. In other words, each cluster has only one subject; the regularity conditions of Gamst et al., 2009 cannot be used directly. At first, regularity conditions to prove the consistency of $\hat{\bld{\theta}}$ are considered. Note that the following proofs are almost the same as Gamst et al., 2009 since the situation is only slightly different.

\begin{theo}{$\phantom{}$}\\
\label{theo1}
Assuming that regularity conditions from {\bf C.1} to {\bf C.4} are satisfied. To achieve the consistency $\hat{\bld{\theta}}\stackrel{P}{\to}\bld{\theta}^{0}$, at least the conditions from {\it 1.} to {\it 3-1} or {\it 3-2.} are additionally need to be satisfied:
\begin{description}
\item[{\it 1}.] $\sigma_{u}$ is a fixed value
\item[{\it 2}.] There is no the intercept term in $\tilde{\bld{x}}_{t}$; $\beta_{0}\equiv0$
\item[{\it 3-1}.] There is no the intercept term in $\tilde{\bld{x}}_{w}$; $\alpha_{0}\equiv0$
\item[{\it 3-2}.] $\rho$ is a fixed value
\end{description}
\end{theo}\noindent
The conditions {\it 1.} and {\it 2.} is necessary to identify the baseline hazard $\lambda_{0}(t)$. In other words, common variabilities in subjects are absorbed into $\lambda_{0}(t)$ since it is not assumed to any models. Note that {\it 2.} is the same condition as Gamst et al., 2009; {\it 1.} is affected by difficulty that there is only one subject in each cluster. From here, we additionally assume $\sigma_{u}=1$. The condition {\it 3.} is necessary to identify the parameters for the treatment model. The situation where {\it 3-2}. is assumed is that the features of the unmeasured covariates distribution are completely recognized; this is not realistic. Therefore, we implicitly assume that {\it 3-1}. is satisfied. Next, regularity conditions to prove the asymptotic normality of $\hat{\bld{\theta}}$ are considered.
\begin{theo}{$\phantom{}$}\\
\label{theo2}
Assuming that regularity conditions from {\bf C.1} to {\bf C.4} are satisfied. To achieve the asymptotic normality $\sqrt{n}(\hat{\bld{\theta}}-\bld{\theta}^{0})\stackrel{L}{\to}N(\bld{0},\Sigma),\ \Sigma>O$, at least the same conditions in {\bf Theorem \ref{theo1}.} are additionally need to be satisfied.
\end{theo}\noindent
Since the proof is also the same flow as Theorem 2 of Gamst et al., 2009, it is omitted in this manuscript. From {\bf Theorem \ref{theo2}.}, any additional assumptions are not necessary to achieve the $\sqrt{n}$-consistency; this is the same result as Gamst et al., 2009. 

From {\bf Theorem \ref{theo1}} and {\bf \ref{theo2}}, properties of the nonparametric MLE $\hat{\bld{\theta}}$ have been appeared, however, specific construction of the observed log-likelihood of (\ref{m_ll}) has not been explained yet. In the following subsection, we introduce the construction of the observed log-likelihoods using the EM algorithm.

\subsection{Construction of EM algorithm}
To construct the EM algorithm for $\hat{\bld{\theta}}$, we need to consider the following expectation:
\begin{align}
\label{for2_4}
{\rm E}\left[G(U)|t,\delta,w,\bld{z},\bld{x}\right]=\int G(u)f(u|t,\delta,w,\bld{z},\bld{x})du
\end{align}
where $G(u)$ denote an arbitrary measurable function. Under {\bf Assumption \ref{assu1}}, the following proposition holds.
\begin{prop}{$\phantom{}$}\\
\label{prop1}
Regarding (\ref{for2_4}), the following equations can be obtained when Assumption \ref{assu1} holds:
\begin{itemize}
\item When $w=1$,
$$
{\rm E}\left[G(U)|\tau,\delta,w=1,\bld{z},\bld{x}\right]=\frac{\int G(u)\left(f(t|u,w=1,\bld{x};\bld{\beta})\right)^{\delta}\left(S_{t|w=1,u,x}(t;\bld{\beta})\right)^{1-\delta}\int_{-\tilde{\bld{x}}_{w}^{\top}\bld{\alpha}}^{\infty}f(u,v;\bld{\xi})dvdu}{\int \left(f(t|u,w=1,\bld{x};\bld{\beta})\right)^{\delta}\left(S_{t|w=1,u,x}(t;\bld{\beta})\right)^{1-\delta}\int_{-\tilde{\bld{x}}_{w}^{\top}\bld{\alpha}}^{\infty}f(u,v;\bld{\xi})dvdu}
$$
\item When $w=0$,
$$
{\rm E}\left[G(U)|\tau,\delta,w=0,\bld{z},\bld{x}\right]=\frac{\int G(u)\left(f(t|u,w=0,\bld{x};\bld{\beta})\right)^{\delta}\left(S_{t|w=0,u,x}(t;\bld{\beta})\right)^{1-\delta}\int^{-\tilde{\bld{x}}_{w}^{\top}\bld{\alpha}}_{-\infty}f(u,v;\bld{\xi})dvdu}{\int \left(f(t|u,w=0,\bld{x};\bld{\beta})\right)^{\delta}\left(S_{t|w=0,u,x}(t;\bld{\beta})\right)^{1-\delta}\int^{-\tilde{\bld{x}}_{w}^{\top}\bld{\alpha}}_{-\infty}f(u,v;\bld{\xi})dvdu}
$$
\end{itemize}
\end{prop}\noindent
From {\bf Proposition \ref{prop1}}, the censor can be completely ignored from the following estimating procedures under {\bf Assumption \ref{assu1}}; the censoring model need not be considered.

\subsubsection{E-step}
The $k$-th iterated parameters denote $\hat{\bld{\vartheta}}_{k}$. Note that $\bld{\vartheta}=\left(\bld{\theta}^{\top},(\lambda_{1},\dots,\lambda_{n_{s}})\right)^{\top}$, where $\lambda_{i}=\lambda(t_{i})$ and $n_{s}=\sum_{i=1}^{n}\delta_{i}$. Hence, $\lambda_{i},\ i=1,2,\dots,n$ are considered as parameter; this is the same consideration as Vaida and Xu, 2000. By considering $\lambda_{i}$ as parameters, the profile likelihood approach (c.f. Johansen, 1983 and Klein and Moeschberger, 2006) can be applied. To clarify, an expectation under parameters $\hat{\bld{\vartheta}}_{k}$ is denoted as ${\rm E}_{\hat{\bld{\vartheta}}_{k}}\left[\cdot\right]$. Then, the $k+1$-th E-step is described as follows:
\begin{align}
\label{for1_11}
\ell^{o}_{k+1}(\bld{\vartheta})&=\sum_{i=1}^{n}\delta_{i}\log\lambda(t_{i})+\sum_{i=1}^{n}\delta_{i}\tilde{\bld{x}}_{t,i}^{\top}\bld{\beta}+\sum_{i=1}^{n}\delta_{i}{\rm E}_{\hat{\bld{\vartheta}}_{k}}\left[U|\tau_{i},\delta_{i},w_{i},\bld{z}_{i},\bld{x}_{i}\right]\nonumber\\
&\hspace{0.5cm}-\sum_{i=1}^{n}\Lambda(t_{i})\exp\left\{\tilde{\bld{x}}_{t,i}^{\top}\bld{\beta}\right\}{\rm E}_{\hat{\bld{\vartheta}}_{k}}\left[\exp\{U\}|\tau_{i},\delta_{i},w_{i},\bld{z}_{i},\bld{x}_{i}\right]\nonumber\\
&\hspace{0.5cm}-\frac{1}{2}\sum_{i=1}^{n}{\rm E}_{\hat{\bld{\vartheta}}_{k}}\left[U^2|\tau_{i},\delta_{i},w_{i},\bld{z}_{i},\bld{x}_{i}\right]\nonumber\\
&\hspace{0.5cm}+\sum_{i=1}^{n_{1}}{\rm E}_{\hat{\bld{\vartheta}}_{k}}\left[\left.\log\Phi\left(\frac{\tilde{\bld{x}}_{w,i}^{\top}\bld{\alpha}+\rho U}{\sqrt{1-\rho^2}}\right)\right|\tau_{i},\delta_{i},w_{i}=1,\bld{z}_{i},\bld{x}_{i}\right]\nonumber\\
&\hspace{0.5cm}+\sum_{i=n_{1}+1}^{n}{\rm E}_{\hat{\bld{\vartheta}}_{k}}\left[\left.\log\left(1-\Phi\left(\frac{\tilde{\bld{x}}_{w,i}^{\top}\bld{\alpha}+\rho U}{\sqrt{1-\rho^2}}\right)\right)\right|\tau_{i},\delta_{i},w_{i}=0,\bld{z}_{i},\bld{x}_{i}\right].
\end{align}
From (\ref{for1_11}), the parameter estimation can be considered in two parts: parameters related to the cox model: $(\bld{\beta},\lambda_{i})$ and the others: $(\bld{\alpha},\sigma_{u},\rho)$ in the following $k+1$-th M-step. We assume that $\ell^{o,1}_{k+1}=\ell^{o,1}_{k+1}(\bld{\beta},\lambda_{1},\dots,\lambda_{n_{s}})$ and $\ell^{o,2}_{k+1}=\ell^{o,2}_{k+1}(\bld{\alpha},\sigma_{u},\rho)$ denote
\begin{align}
\label{2_13}
\ell^{o,1}_{k+1}&=\sum_{i=1}^{n}\delta_{i}\log\lambda(t_{i})+\sum_{i=1}^{n}\delta_{i}\tilde{\bld{x}}_{t,i}^{\top}\bld{\beta}+\sum_{i=1}^{n}\delta_{i}{\rm E}_{\hat{\bld{\vartheta}}_{k}}\left[U|\tau_{i},\delta_{i},w_{i},\bld{z}_{i},\bld{x}_{i}\right]\nonumber\\
&\hspace{0.5cm}-\sum_{i=1}^{n}\Lambda(t_{i})\exp\left\{\tilde{\bld{x}}_{t,i}^{\top}\bld{\beta}\right\}{\rm E}_{\hat{\bld{\vartheta}}_{k}}\left[\exp\{U\}|\tau_{i},\delta_{i},w_{i},\bld{z}_{i},\bld{x}_{i}\right],\nonumber\\
\ell^{o,2}_{k+1}&=-\frac{1}{2}\sum_{i=1}^{n}{\rm E}_{\hat{\bld{\vartheta}}_{k}}\left[U^2|\tau_{i},\delta_{i},w_{i},\bld{z}_{i},\bld{x}_{i}\right]\nonumber\\
&\hspace{0.5cm}+\sum_{i=1}^{n_{1}}{\rm E}_{\hat{\bld{\vartheta}}_{k}}\left[\left.\log\Phi\left(\frac{\tilde{\bld{x}}_{w,i}^{\top}\bld{\alpha}+\rho U}{\sqrt{1-\rho^2}}\right)\right|\tau_{i},\delta_{i},w_{i}=1,\bld{z}_{i},\bld{x}_{i}\right]\nonumber\\
&\hspace{0.5cm}+\sum_{i=n_{1}+1}^{n}{\rm E}_{\hat{\bld{\vartheta}}_{k}}\left[\left.\log\left(1-\Phi\left(\frac{\tilde{\bld{x}}_{w,i}^{\top}\bld{\alpha}+\rho U}{\sqrt{1-\rho^2}}\right)\right)\right|\tau_{i},\delta_{i},w_{i}=0,\bld{z}_{i},\bld{x}_{i}\right],
\end{align}
respectively. Therefore, $\ell^{o}_{k+1}=\ell^{o,1}_{k+1}+\ell^{o,2}_{k+1}$. When estimating the expectation
$$
{\rm E}_{\hat{\bld{\vartheta}}_{k}}\left[G(U)|\tau,\delta,w,\bld{z},\bld{x}\right],
$$
following the below procedures:
\begin{enumerate}
\item {\it Generating monte carlo samples}
$$
u_{1},u_{2},\dots,u_{B}\sim N(0,\hat{\sigma}_{u,k}^2)
$$
\item {\it Monte carlo integration}
\begin{itemize}
\item When $w=1$,
\begin{align*}
\widehat{\rm E}\left[G(U)|\tau,\delta,w=1,\bld{z},\bld{x}\right]&\\
&\hspace{-4cm}=\frac{\sum_{b=1}^{B}G(u_{b})\left(f(t|u_{b},w=1,\bld{x};\bld{\beta})\right)^{\delta}\left(S_{t|w=1,u_{b},x}(t;\bld{\beta})\right)^{1-\delta}\Phi\left(\frac{\tilde{\bld{x}}_{w,i}^{\top}\bld{\alpha}+\rho u_{b}}{\sqrt{1-\rho^2}}\right)}{\sum_{b=1}^{B}\left(f(t|u_{b},w=1,\bld{x};\bld{\beta})\right)^{\delta}\left(S_{t|w=1,u_{b},x}(t;\bld{\beta})\right)^{1-\delta}\Phi\left(\frac{\tilde{\bld{x}}_{w,i}^{\top}\bld{\alpha}+\rho u_{b}}{\sqrt{1-\rho^2}}\right)}
\end{align*}
\item When $w=0$,
\begin{align*}
\widehat{\rm E}\left[G(U)|\tau,\delta,w=0,\bld{z},\bld{x}\right]&\\
&\hspace{-4cm}=\frac{\sum_{b=1}^{B}G(u_{b})\left(f(t|u_{b},w=0,\bld{x};\bld{\beta})\right)^{\delta}\left(S_{t|w=0,u_{b},x}(t;\bld{\beta})\right)^{1-\delta}\left(1-\Phi\left(\frac{\tilde{\bld{x}}_{w,i}^{\top}\bld{\alpha}+\rho u_{b}}{\sqrt{1-\rho^2}}\right)\right)}{\sum_{b=1}^{B}\left(f(t|u_{b},w=0,\bld{x};\bld{\beta})\right)^{\delta}\left(S_{t|w=0,u_{b},x}(t;\bld{\beta})\right)^{1-\delta}\left(1-\Phi\left(\frac{\tilde{\bld{x}}_{w,i}^{\top}\bld{\alpha}+\rho u_{b}}{\sqrt{1-\rho^2}}\right)\right)}
\end{align*}
\end{itemize}
\end{enumerate}
From the above, the observed likelihoods $\ell^{o,1}_{k+1}$ and $\ell^{o,2}_{k+1}$ can be constructed.
\subsubsection{M-step}
By the same approach as Vaida and Xu, 2000, $\ell^{o,1}_{k+1}$ becomes maximize when
$$
\hat{\lambda}_{i,k+1}=\frac{1}{\sum_{i\leq j}\exp\left\{\tilde{\bld{x}}_{t,i}^{\top}\bld{\beta}+\log\left(\widehat{\rm E}_{\hat{\bld{\vartheta}}_{k}}\left[\exp\{U\}|\tau_{j},\delta_{j},w_{j},\bld{z}_{j},\bld{x}_{j}\right]\right)\right\}}.
$$
Then, the profile log-likelihood becomes
\begin{align}
\label{for1_12}
\sum_{i=1}^{n}\delta_{i}\left(\tilde{\bld{x}}_{t,i}^{\top}\bld{\beta}-\log\sum_{i\leq j}\exp\left\{\tilde{\bld{x}}_{t,i}^{\top}\bld{\beta}+\log\left(\widehat{\rm E}_{\hat{\bld{\vartheta}}_{k}}\left[\exp\{U\}|\tau_{j},\delta_{j},w_{j},\bld{z}_{j},\bld{x}_{j}\right]\right)\right\}\right).
\end{align}
Maximizing (\ref{for1_12}), $\ell^{o,1}_{k+1}$ becomes maximum:
$$
\hat{\bld{\beta}}_{k+1}=\argmax_{\bld{\beta}}\ell^{o,1}_{k+1}(\bld{\beta},\hat{\lambda}_{1,k+1},\dots,\hat{\lambda}_{n_{s},k+1}).
$$
Whereas, $\ell^{o,2}_{k+1}$ can be maximized simply:
$$
(\hat{\bld{\alpha}}_{k+1},\hat{\rho}_{k+1})=\argmax_{\bld{\alpha},\rho}\ell^{o,2}_{k+1}(\bld{\alpha},\rho).
$$
From the above, the $k+1$-th parameter estimator $\hat{\bld{\vartheta}}_{k+1}$ can be obtained. When
$$
\left|\left|\hat{\bld{\vartheta}}_{k+1}-\hat{\bld{\vartheta}}_{k}\right|\right|<\varepsilon,
$$
the parameter estimator becomes convergence sufficiently; the parameter estimator $\hat{\bld{\vartheta}}=\hat{\bld{\vartheta}}_{k+1}$, where $\varepsilon>0$.
\section{Simulations}
In this section, we confirm properties of our proposed method under some situations comparing with 1) an ordinary partial likelihood estimator except for unmeasured covariates and 2) an ordinary partial likelihood estimator including unmeasured covariates. Note that the estimator 1) may have some biases, and the estimator 2) is an infeasible estimator since we cannot observe unmeasured covatiates. A part of simulation settings refer to Kianian et al., 2019. The number of iterations for all simulations are 1,000.
\subsection{Valid unmeasured covariate and treatment model situation}
At first, we confirm performances of our proposed estimator under a valid model situation. Under the situation, simulation settings are as follows:
\begin{description}
\item{{\bf Covariate}}
$$
X_{i}\stackrel{i.i.d.}{\sim} Unif(-1,1)
$$
\item{{\bf Unmeasured covariates}}
$$
\left(
\begin{array}{c}
V_{i}\\
U_{i}
\end{array}
\right)\stackrel{i.i.d.}{\sim} N_{2}\left(\bld{0}_{2},
\left(
\begin{array}{cc}
1&\sigma_{uv}\\
&\sigma_{u}^2
\end{array}
\right)
\right)
$$
\begin{itemize}
\item High variance: $\sigma_{u}^2=1$
\item Low variance: $\sigma_{u}^2=0.1$
\end{itemize}
\begin{itemize}
\item Strong correlation: $\sigma_{uv}=0.4\times \sigma_{u}$
\item Weak correlation: $\sigma_{uv}=0.1\times \sigma_{u}$
\end{itemize}
\item{{\bf Instrumental variable}}
$$
Z_{i}\stackrel{i.i.d.}{\sim} Gamma(2,2)
$$
\item{{\bf Treatment: Probit model}}
$$
W_{i}=\bld{1}\left\{\alpha_{wz}\times Z_{i}+V_{i}\geq0\right\}
$$
\begin{itemize}
\item Strong IV: $\alpha_{wz}=1$
\item Weak IV: $\alpha_{wz}=0.1$
\end{itemize}
\item{{\bf Hazard model}}
$$
\lambda(t_{i})=\exp\left\{0.5\times W_{i}+0.2\times X_{i}+U_{i}\right\}
$$
\item{{\bf Censor}}
$$
C_{i}\stackrel{i.i.d.}{\sim} \exp\{0.5\}
$$
\end{description}
Our interest is treatment effects 0.5 in the hazard model, and how accurately estimating treatment effects. Under the setting, the hazard ratio related to $W_{i}$ becomes $\exp\{0.5\}=1.649$. Note that a proportion of censoring is approximately 70\%. Also, the number of the Monte Carlo sampling ($B$) is as follows:
\begin{itemize}
\item When a small sample situation ($n=200$), $B=40$
\item When a large sample situation ($n=500$), $B=100$
\end{itemize}
Under the situation, we confirm four scenarios (see table \ref{tab0}):

\begin{table}[h]
\begin{center}
\caption{Summary of scenarios}
\begin{tabular}{|c|c|}\hline
{\bf Scenario \#}&{\bf Scenario}\\\hline
\# 1&\begin{tabular}{c}
High variance \& strong correlation of\\
unmeasured covariates and strong IV
\end{tabular}\\\hline
\# 2&\begin{tabular}{c}
Low variance \& strong correlation of\\
unmeasured covariates and strong IV
\end{tabular}\\\hline
\# 3&\begin{tabular}{c}
High variance \& weak correlation of\\
unmeasured covariates and strong IV
\end{tabular}\\\hline
\# 4&\begin{tabular}{c}
High variance \& strong correlation of\\
unmeasured covariates and weak IV
\end{tabular}\\\hline
\end{tabular}
\label{tab0}
\end{center}
\end{table}\noindent
Scenario \#1 means the reference of our simulations. In each scenario (from \#2 to \#4), we change each parameter and confirm the variation of estimates of the hazard ratio. Summaries of each estimator for the hazard ratio are as follows (see table \ref{tab1}):

\begin{table}[h]
\begin{center}
\caption{Summary of estimators for the hazard ratio}
\scalebox{0.8}{
\begin{tabular}{|c|c||c|c|c|c|c|c|}\hline
{{\bf Scenario \#}}&{{\bf Method}}&\multicolumn{3}{|c|}{Small sample $n=200$}&\multicolumn{3}{|c|}{Large sample $n=500$}\\\cline{3-8}
&&{Mean(SD)}&{Median(Range)}&RMSE&{Mean(SD)}&{Median(Range)}&RMSE\\\hline
\# 1&{{\bf Proposed}}&1.901(0.738)&1.738(0.63-5.73)&0.780&1.775(0.423)&1.712(0.84-4.44)&0.441\\\cline{2-8}
&{{\bf Ordinary}}&2.291(0.593)&2.193(1.06-5.54)&0.874&2.218(0.340)&2.190(1.39-4.00)&0.663\\\cline{2-8}
&
\begin{tabular}{c}
{\bf Ordinary}\\
{\bf (infeasible)}
\end{tabular}
&1.728(0.456)&1.671(0.82-3.94)&0.463&1.673(0.267)&1.648(0.98-3.04)&0.268\\\hline
\# 2&{{\bf Proposed}}&1.723(0.639)&1.595(0.62-5.99)&0.644&1.619(0.371)&1.564(0.70-3.23)&0.372\\\cline{2-8}
&{{\bf Ordinary}}&1.969(0.446)&1.920(1.00-3.94)&0.549&1.935(0.276)&1.904(1.24-2.92)&0.398\\\cline{2-8}
&\begin{tabular}{c}
{\bf Ordinary}\\
{\bf (infeasible)}
\end{tabular}&1.687(0.391)&1.648(0.79-3.36)&0.392&1.662(0.246)&1.636(1.09-2.68)&0.246\\\hline
\# 3&{{\bf Proposed}}&1.531(0.588)&1.425(0.53-5.43)&0.600&1.436(0.312)&1.394(0.77-2.97)&0.377\\\cline{2-8}
&{{\bf Ordinary}}&1.615(0.369)&1.574(0.87-3.81)&0.371&1.574(0.217)&1.561(1.00-2.41)&0.230\\\cline{2-8}
&\begin{tabular}{c}
{\bf Ordinary}\\
{\bf (infeasible)}
\end{tabular}&1.707(0.412)&1.654(0.79-3.54)&0.417&1.662(0.239)&1.645(1.01-2.58)&0.239\\\hline
\# 4&{{\bf Proposed}}&1.840(0.552)&1.768(0.82-5.35)&0.584&1.750(0.298)&1.719(1.06-3.14)&0.315\\\cline{2-8}
&{{\bf Ordinary}}&2.283(0.424)&2.236(1.20-4.76)&0.763&2.256(0.261)&2.227(1.57-3.51)&0.661\\\cline{2-8}
&\begin{tabular}{c}
{\bf Ordinary}\\
{\bf (infeasible)}
\end{tabular}&1.679(0.334)&1.650(0.92-3.69)&0.335&1.656(0.201)&1.636(1.12-2.58)&0.201\\\hline
\end{tabular}
}
\label{tab1}
\end{center}
\end{table}\noindent
In scenario \#1, our proposed estimator can be estimated with some accuracy in small sample, whereas, an ordinary estimator obviously has some biases. In large sample, not only the SD but also the RMSE of our proposed estimator decreases; we can confirm that our proposed estimator has the consistency through simulation. In scenario \#2, although our estimator has underestimates, the results are almost identical to scenario \#1. Note that the bias of an ordinary estimator is less than scenario \#1 since an unmeasured covariate has the low variance; the effect of an unmeasured covariate on survival times is small. In scenario \#3, we confirm through additional simulation results later, our estimator has obviously underestimates in both small and large sample sample. However, an ordinary estimator has the less bias; this is because the effect of an unmeasured covariates has a small effect on survival times also. In scenario \#4, regardless of the strength of an instrument variable, our proposed estimator can be estimated accuracy. Through the four scenarios, we can confirm that the accuracy of our proposed estimator is related to the correlation of unmeasured covariates, not the strength of instrument variables. 

As we mentioned previously, our estimator has a serious bias in weak correlation situation. Summaries of estimators are as follows (see table \ref{tab2}):

\begin{table}[h]
\begin{center}
\caption{Summary of estimators about scenario \# 3}
\scalebox{0.9}{
\begin{tabular}{|c|c|c||c|c|c|c|}\hline
{{\bf Scenario \#}}&{{\bf Method}}&{{\bf Parameter}}&\multicolumn{4}{|c|}{Large sample $n=500$}\\\cline{4-7}
&&&{Mean(SD)}&{Median(Range)}&RMSE&\begin{tabular}{c}
Coefficient of\\
Variation (CV)
\end{tabular}\\\hline
\# 3&{{\bf Proposed}}&$\beta_{t}$&0.343(0.217)&0.336(-0.48-1.01)&0.267&0.633\\\cline{3-7}
&&$\beta_{x}$&0.179(0.129)&0.180(-0.25-0.61)&0.130&0.721\\\cline{3-7}
&&$\alpha$&1.048(0.082)&1.042(0.80-1.35)&0.095&0.078\\\cline{3-7}
&&$\sigma_{u}$&1.151(0.016)&1.151(1.11-1.21)&0.152&0.014\\\cline{3-7}
&&$\rho$&0.187(0.095)&0.192(-0.12-0.48)&0.129&0.508\\\hline
\end{tabular}
}
\label{tab2}
\end{center}
\end{table}\noindent
Parameters related to the hazard model have some biases and CVs, however, $\alpha$ and $\sigma_{u}$ have small ones. Remark that the estimate of $\rho$ also have 1) some biases, 2) a relatively large RMSE, and 3) a large CV. This ma be the similar result as the variance of an ordinary sample correlation (Fisher, 1921):
$$
Var(r)\approx\frac{1-r^2}{\sqrt{n}},
$$
where $r$ is an ordinary sample correlation; the smaller a sample correlation, the larger the variance of the estimator. Therefore, the variability of $\beta_{w},\, \beta_{x}$ may be related to the variability of $\rho$. Thorough the additional simulation, we confirm that not only estimates of hazard models but also correlations of unmeasured covariates have some biases and large variation when the correlation is small. Unfortunately the situation cannot be confirmed by using observed data; therefore some sensitivity analyses are necessary.

\subsection{Some invalid unmeasured covariate or treatment model situation}
Next, we confirm performances of our proposed estimator under some invalid model situation. Many settings are the same as the previous subsection, but we use the invalid unmeasured covariate and treatment models as follows:
\begin{description}
\item{{\bf Treatment: Logistic model}}
$$
W_{i}=\bld{1}\left\{Z_{i}+V_{i}\geq0\right\},\ \ V_{i}\stackrel{i.i.d.}{\sim}Logistic(0,1)
$$
$$
U_{i}=0.45\times V_{i}+\varepsilon_{i},\ \ \varepsilon_{i}\stackrel{i.i.d.}{\sim}N(0,1)
$$
\item{{\bf Unmeasured covariates: Asymmetric distribution}}\\
$$
U_{i}=U'_{i}-\bar{U}',\ \ U'_{i}\sim Gamma(\nu(V_{i}),1),\ \nu(V_{i})=0.15\times \exp\left\{V_{i}\right\}
$$
\item{{\bf Unmeasured covariates: Heavy tail distribution}}\\
We use the linear Regression Under Heavy-Tailed Distributions (see Lange and Sinsheimer, 1993 and the description of heavyLm (R function)).
$$
U_{i}\sim t(\mu(V_{i}),4),\ \mu(V_{i})=0.5\times V_{i}
$$
\end{description}
Note that $Cor(U,V)\approx 0.4$ in the above settings. The relationship between the above settings and scenario numbers is as follows (see table \ref{tab4}):

\begin{table}[h]
\begin{center}
\caption{Summary of scenarios}
\begin{tabular}{|c|c|}\hline
{\bf Scenario \#}&{\bf Scenario}\\\hline
\# 5&\begin{tabular}{c}
Treatment: Logistic model
\end{tabular}\\\hline
\# 6&\begin{tabular}{c}
Unmeasured covariates: Asymmetric distribution
\end{tabular}\\\hline
\# 7&\begin{tabular}{c}
Unmeasured covariates: Heavy tail distribution
\end{tabular}\\\hline
\end{tabular}
\label{tab4}
\end{center}
\end{table}\noindent
Summaries of each estimator for a hazard ratio are as follows (see table \ref{tab5}):

\begin{table}[h]
\begin{center}
\caption{Summary of estimators for a hazard ratio}
\begin{tabular}{|c|c||c|c|c|}\hline
{{\bf Scenario \#}}&{{\bf Method}}&\multicolumn{3}{|c|}{Large sample $n=500$}\\\cline{3-5}
&&{Mean(SD)}&{Median(Range)}&RMSE\\\hline
\# 5&{{\bf Proposed}}&1.702(0.387)&1.638(0.89-3.65)&0.390\\\cline{2-5}
&{{\bf Ordinary}}&1.907(0.258)&1.887(1.29-2.89)&0.365\\\cline{2-5}
&\begin{tabular}{c}
{\bf Ordinary}\\
{\bf (infeasible)}
\end{tabular}&1.668(0.232)&1.648(1.10-2.59)&0.233\\\hline
\# 6&{{\bf Proposed}}&1.903(0.470)&1.807(0.99-4.78)&0.535\\\cline{2-5}
&{{\bf Ordinary}}&2.350(0.352)&2.307(1.45-3.89)&0.784\\\cline{2-5}
&
\begin{tabular}{c}
{\bf Ordinary}\\
{\bf (infeasible)}
\end{tabular}
&1.664(0.260)&1.637(1.01-2.65)&0.261\\\hline
\# 7&{{\bf Proposed}}&1.782(0.412)&1.727(0.93-3.90)&0.433\\\cline{2-5}
&{{\bf Ordinary}}&2.244(0.333)&2.210(1.47-4.29)&0.682\\\cline{2-5}
&\begin{tabular}{c}
{\bf Ordinary}\\
{\bf (infeasible)}
\end{tabular}&1.659(0.254)&1.643(1.05-3.17)&0.254\\\hline
\end{tabular}
\label{tab5}
\end{center}
\end{table}\noindent
In scenario \# 5, there are only small biases, and it is founded that the effect of model misspecification related to treatment is limited. This may be derived from the similarity of Normal distribution and Logistic distribution. In scenario \# 6, there are some biases, however, the impact is small compared to ordinary estimates. In scenario \# 7, we think the scenario is the most notable, even if an unmeasured covariate has heavy tail, our proposed estimator has the small RMSE to some extent. From scenario \#6 and \#7, it is founded that the effect of model misspecification related to unmeasured covariates is limited. Additionally, the effect of unmeasured ``outliers"  is also limited; this is one of the important property of our proposed estimator. Therefore, we conclude that our proposed estimator has the robust property in the sense of deriving the valid estimates.

\section{Conclusions and Future Works}
In this paper, we propose the new estimator to overcome the unmeasured covariates problem when there are binary treatments and multiple IVs. By applying the idea of a frailty to unmeasured covariates, we can construct the EM algorithm, and the derived estimator has the consistency; these are the same approach and result as Vaida and Xu, 2000. Through some simulations, we can confirm that our estimator has good performances except the situation where there are only small correlation between unmeasured variables. Even if we misspecify the treatment model or the unmeasured covariate model, our proposed estimator has the robust property in the sense of deriving the valid estimates. Note that when an unmeasured covariate has heavy tail, the robustness is not violated.

Although we can confirm the robustness, the possibility of some bias cannot be denied when (\ref{for1_1}) is misspecified. Additionally, the logistic model is more often used than the probit model for a treatment model in biometrics and related fields. To relax the assumption, the copula methods may be applied (Nelsen, 2007 and Emura et al., 2017). By using a copula to unmeasured variables, we only assume marginal models related to a treatment model and a hazard model separately. However, we cannot apply the proposed procedure in this paper simply; we need to use more complicated estimating procedures such as the MCMC. Also, we need to expand our proposed method applying to various outcome models. For example, expanding to  the competing risk model (Ying et al., 2019) or semi-competing model (Fine et al., 2001) may also be important for applications. 

\newpage
\begin{singlespace}

\end{singlespace}

\newpage
\appendix
\section{Regularity conditions}
These are the same conditions as Gamst et al., 2009.
\begin{description}
\item[C.1] Let $\tau>0$. There is some $\varepsilon>0$ such that $P(C\geq\tau|w,\bld{z},\bld{x})\geq\varepsilon$, almost surely.
\item[C.2] The baseline hazard function $\lambda_{0}(t)>0$ and is continuous on the finite time interval $[0,\tau]$. Note that $[0,\tau]\subset \mathcal{T}$
\item[C.3] $\mathcal{X}$ and $\mathcal{W}$ are bounded. Note that the boundedness of $\mathcal{W}$ is hold clearly when the treatment $W$ is dichotomous.
\item[C.4] The true parameters $\bld{\theta}^{0}$ is an element of the interior of a known compact set $\Theta$.
\end{description}

\section{Calculations and Proofs}
\subsection{Expansion of (\ref{for2_3})}
We show only the situation where $w=1$. When $w=1$, $\bld{z}$, and $\bld{x}$ are fixed, $v\geq -\tilde{\bld{x}}_{w}^{\top}\bld{\alpha}$ from (\ref{for1_2}). Then, 
\begin{align*}
f(u|w=1,\bld{z},\bld{x})&=f(u|v\geq-\tilde{\bld{x}}_{w}^{\top}\bld{\alpha},\bld{z},\bld{x})\\
&=\frac{f(u,v\geq-\tilde{\bld{x}}_{w}^{\top}\bld{\alpha}|\bld{z},\bld{x})}{f(v\geq-\tilde{\bld{x}}_{w}^{\top}\bld{\alpha}|\bld{z},\bld{x})}=\frac{\int_{-\tilde{\bld{x}}_{w}^{\top}\bld{\alpha}}^{\infty}f(u,v;\bld{\xi})dv}{f(w=1|\bld{z},\bld{x};\bld{\alpha})}
\end{align*}
Therefore, (\ref{for2_3}) is obtained.

\subsection{Proof of Theorem \ref{theo1}}
The proof is the same flow as Theorem 1 of Gamst et al., 2009 except for the parameter identification part (from the second half of page 5 to page 6 of the supplementary material). Therefore, it is only necessary to consider the parameter identification in this proof. The following equation is only considered:
\begin{align}
\label{app1_2}
\int_{u}\prod_{i=1}^{k}\lambda^{*}(0)\exp\left\{\tilde{\bld{x}}_{t,i}^{\top}\bld{\beta}^{*}+u\right\}\left(\int_{-\tilde{\bld{x}}_{w,i}^{\top}\bld{\alpha}^{*}}^{\infty}f(u_{i},v;\bld{\xi}^{*})dv\right)^{w_{i}}\left(\int^{-\tilde{\bld{x}}_{w,i}^{\top}\bld{\alpha}^{*}}_{-\infty}f(u_{i},v;\bld{\xi}^{*})dv\right)^{1-w_{i}}du&\nonumber\\
&\hspace{-15cm}=\int_{u}\prod_{i=1}^{k}\lambda^{0}(0)\exp\left\{\tilde{\bld{x}}_{t,i}^{\top}\bld{\beta}^{0}+u\right\}\left(\int_{-\tilde{\bld{x}}_{w,i}^{\top}\bld{\alpha}^{0}}^{\infty}f(u_{i},v;\bld{\xi}^{0})dv\right)^{w_{i}}\left(\int^{-\tilde{\bld{x}}_{w,i}^{\top}\bld{\alpha}^{0}}_{-\infty}f(u_{i},v;\bld{\xi}^{0})dv\right)^{1-w_{i}}du
\end{align}
Note that $k$ denotes the number of subjects who have events at $t=0$. This is from the regularity condition {\bf C.2}. We only consider the situation where $w=1$. The integration for $u$ of (\ref{app1_2}) becomes
\begin{align}
\label{app1_3}
\int_{u}\exp\left\{u\right\}\left[\int_{-\tilde{\bld{x}}_{w,i}^{\top}\bld{\alpha}}^{\infty}\frac{1}{\sqrt{2\pi(1-\rho^2)}}\exp\left\{-\frac{\left(v-\frac{\rho}{\sigma_{u}}u\right)^2}{2(1-\rho^2)}\right\}dv\right]\frac{1}{\sqrt{2\pi\sigma_{u}^2}}\exp\left\{-\frac{u^2}{2\sigma_{u}^2}\right\}du&\nonumber\\
&\hspace{-14cm}=\frac{1}{2\pi\sqrt{(1-\rho^2)\sigma_{u}^2}}\int_{-\tilde{\bld{x}}_{w,i}^{\top}\bld{\alpha}}^{\infty}\int_{u}\exp\left\{u-\frac{\left(v-\frac{\rho}{\sigma_{u}}u\right)^2}{2(1-\rho^2)}-\frac{u^2}{2\sigma_{u}^2}\right\}dudv.
\end{align}
For (\ref{app1_3}), the term in $\left\{\cdot\right\}$ is
\begin{align*}
u-\frac{\left(v-\frac{\rho}{\sigma_{u}}u\right)^2}{2(1-\rho^2)}-\frac{u^2}{2\sigma_{u}^2}&=\frac{1}{2(1-\rho^2)\sigma^{2}_{u}}\left\{(u-\sigma_{u}(\sigma_{u}(1-\rho^2)+\rho v))^2-\sigma_{u}^2(1-\rho^2)(v-\sigma_{u}\rho)^2+(1-\rho^2)\sigma_{u}^4\right\}\\
&=\frac{1}{2(1-\rho^2)\sigma^{2}_{u}}\left\{(u-\sigma_{u}(\sigma_{u}(1-\rho^2)+\rho v))^2\right\}-\frac{(v-\sigma_{u}\rho)^2}{2}+\frac{\sigma_{u}^2}{2}.
\end{align*}
Therefore, $u$ is regarded as $U\sim N(\sigma_{u}(\sigma_{u}(1-\rho^2)+\rho v),(1-\rho^2)\sigma^{2}_{u})$. Then, (\ref{app1_3}) becomes
$$
\exp\left\{\frac{\sigma_{u}^2}{2}\right\}\frac{1}{\sqrt{2\pi}}\int_{-\tilde{\bld{x}}_{w,i}^{\top}\bld{\alpha}}^{\infty}\exp\left\{-\frac{(v-\sigma_{u}\rho)^2}{2}\right\}dv=\exp\left\{\frac{\sigma_{u}^2}{2}\right\}\Phi\left(\tilde{\bld{x}}_{w,i}^{\top}\bld{\alpha}+\sigma_{u}\rho\right).
$$
Therefore, (\ref{app1_2}) becomes
\begin{align}
\label{app1_4}
\prod_{i=1}^{k}\lambda^{*}(0)\exp\left\{\tilde{\bld{x}}_{t,i}^{\top}\bld{\beta}^{*}+\frac{{\sigma_{u}^{*}}^2}{2}\right\}\left(\Phi\left(\tilde{\bld{x}}_{w,i}^{\top}\bld{\alpha}^{*}+\sigma_{u}^{*}\rho^{*}\right)\right)^{w_{i}}\left(1-\Phi\left(\tilde{\bld{x}}_{w,i}^{\top}\bld{\alpha}^{*}+\sigma_{u}^{*}\rho^{*}\right)\right)^{1-w_{i}}&\nonumber\\
&\hspace{-12cm}=\prod_{i=1}^{k}\lambda^{0}(0)\exp\left\{\tilde{\bld{x}}_{t,i}^{\top}\bld{\beta}^{0}+\frac{{\sigma_{u}^{0}}^2}{2}\right\}\left(\Phi\left(\tilde{\bld{x}}_{w,i}^{\top}\bld{\alpha}^{0}+\sigma_{u}^{0}\rho^{0}\right)\right)^{w_{i}}\left(1-\Phi\left(\tilde{\bld{x}}_{w,i}^{\top}\bld{\alpha}^{0}+\sigma_{u}^{0}\rho^{0}\right)\right)^{1-w_{i}}.
\end{align}
From (\ref{app1_4}), a sufficient condition to achieve uniqueness of the true parameters is that $^{\forall}i\in\{1,2,\dots,n\}$,
\begin{align}
\label{app1_5}
\left(\log\left\{\lambda^{*}(0)\right\}-\log\left\{\lambda^{0}(0)\right\}\right)+\left({\sigma_{u}^{0}}^2-{\sigma_{u}^{*}}^2\right)+\tilde{\bld{x}}_{t,i}^{\top}\left(\bld{\beta}^{*}-\bld{\beta}^{0}\right)\nonumber\\
&\hspace{-10cm}
+w_{i}\left(\log\frac{\Phi\left(\tilde{\bld{x}}_{w,i}^{\top}\bld{\alpha}^{*}+\sigma_{u}^{*}\rho^{*}\right)}{\Phi\left(\tilde{\bld{x}}_{w,i}^{\top}\bld{\alpha}^{0}+\sigma_{u}^{0}\rho^{0}\right)}\right)+(1-w_{i})\left(\log\frac{1-\Phi\left(\tilde{\bld{x}}_{w,i}^{\top}\bld{\alpha}^{*}+\sigma_{u}^{*}\rho^{*}\right)}{1-\Phi\left(\tilde{\bld{x}}_{w,i}^{\top}\bld{\alpha}^{0}+\sigma_{u}^{0}\rho^{0}\right)}\right)=0
\end{align}
is hold only when $\bld{\beta}^{*}=\bld{\beta}^{0}$, $\bld{\xi}^{*}=\bld{\xi}^{0}$, and $\lambda^{*}(0)=\lambda^{0}(0)$. Clearly, it is necessary that 1) $\lambda(0)$ or $\sigma_{u}$ is fixed value, and 2) $\tilde{\bld{x}}_{t,i}$ does not include the intercept. For 1), ``$\sigma_{u}$ is fixed value" is more appropriate since we assume the CPHM. It is also necessary that 3) $\rho$ is fixed value, or 4) $\tilde{\bld{x}}_{w,i}$ does not include the intercept. Therefore, the statement of Theorem \ref{theo1} has been proved.

\subsection{Proof of Proposition \ref{prop1}}
Calculating the expectation (\ref{for2_4}):
$$
\int G(u)f(u|t,\delta,w,\bld{z},\bld{x})du=\frac{\int G(u)f(t,\delta,w,u,\bld{z},\bld{x})du}{\int f(t,\delta,w,u,\bld{z},\bld{x})du}
$$
We show only the situation where $w=1$. From (\ref{for2_3}), the above formula becomes
\begin{align*}
\frac{\int G(u)f(t,\delta,w=1,u,\bld{z},\bld{x})du}{\int f(t,\delta,w=1,u,\bld{z},\bld{x})du}&\\
&\hspace{-5.25cm}=\scalebox{1.3}{$\frac{\int G(u)\left(S_{c|w=1,x,z}(t;\bld{\beta})\right)^{\delta}\left(f_{c}(t|w=1,\bld{x},\bld{z})\right)^{1-\delta}\left(f(t|w=1,\bld{x},u;\bld{\beta})\right)^{\delta}\left(S_{t|w=1,x,u}(t;\bld{\beta})\right)^{1-\delta}\int_{-\tilde{\bld{x}}_{w}^{\top}\bld{\alpha}}^{\infty}f(u,v;\bld{\xi})dvdu}{\int \left(S_{c|w=1,x,z}(t;\bld{\beta})\right)^{\delta}\left(f_{c}(t|w=1,\bld{x},\bld{z})\right)^{1-\delta}\left(f(t|w=1,\bld{x},u;\bld{\beta})\right)^{\delta}\left(S_{t|w=1,x,u}(t;\bld{\beta})\right)^{1-\delta}\int_{-\tilde{\bld{x}}_{w}^{\top}\bld{\alpha}}^{\infty}f(u,v;\bld{\xi})dvdu}$}\\
&\hspace{-5.25cm}=\scalebox{1.3}{$\frac{\left(S_{c|w=1,x,z}(t;\bld{\beta})\right)^{\delta}\left(f_{c}(t|w=1,\bld{x},\bld{z})\right)^{1-\delta}\int G(u)\left(f(t|w=1,\bld{x},u;\bld{\beta})\right)^{\delta}\left(S_{t|w=1,x,u}(t;\bld{\beta})\right)^{1-\delta}\int_{-\tilde{\bld{x}}_{w}^{\top}\bld{\alpha}}^{\infty}f(u,v;\bld{\xi})dvdu}{ \left(S_{c|w=1,x,z}(t;\bld{\beta})\right)^{\delta}\left(f_{c}(t|w=1,\bld{x},\bld{z})\right)^{1-\delta}\int \left(f(t|w=1,\bld{x},u;\bld{\beta})\right)^{\delta}\left(S_{t|w=1,x,u}(t;\bld{\beta})\right)^{1-\delta}\int_{-\tilde{\bld{x}}_{w}^{\top}\bld{\alpha}}^{\infty}f(u,v;\bld{\xi})dvdu}$}\\
&\hspace{-5.25cm}=\scalebox{1.3}{$\frac{\int G(u)\left(f(t|w=1,\bld{x},u;\bld{\beta})\right)^{\delta}\left(S_{t|w=1,x,u}(t;\bld{\beta})\right)^{1-\delta}\int_{-\tilde{\bld{x}}_{w}^{\top}\bld{\alpha}}^{\infty}f(u,v;\bld{\xi})dvdu}{\int \left(f(t|w=1,\bld{x},u;\bld{\beta})\right)^{\delta}\left(S_{t|w=1,x,u}(t;\bld{\beta})\right)^{1-\delta}\int_{-\tilde{\bld{x}}_{w}^{\top}\bld{\alpha}}^{\infty}f(u,v;\bld{\xi})dvdu}$}
\end{align*}
As a result, the {\bf Proposition \ref{prop1}} is obtained.

\end{document}